\documentclass[aip, cha, reprint]{revtex4-1}
\usepackage{xspace,units}
\usepackage{graphicx, multirow}
\usepackage{amssymb, amsmath, amsfonts}
\usepackage{dcolumn}% Align table columns on decimal point
\usepackage{bm}% bold math
\usepackage{enumitem}

\usepackage[utf8]{inputenc}
\usepackage[T1]{fontenc}

\newcommand {\CEE}{$\mathcal{C}^{\mathrm{E}}_{\mathrm{e}}$\xspace}

\newcommand {\CBE}{$\mathcal{C}^{\mathrm{B}}_{\mathrm{e}}$\xspace}

\newcommand{\defi}{\mathrel{\mathop:}=}

\newcommand{\iu}{{i\mkern1mu}}
\newcommand {\bfA} {\ensuremath{\boldsymbol{A}}\xspace}
\newcommand {\bftA} {\ensuremath{\tilde{\boldsymbol{A}}}\xspace}
\newcommand {\cpl} {\ensuremath{K}\xspace}
\newcommand {\NWebs}{\ensuremath{N_W}\xspace}
\newcommand {\topoA}{$\mathrm{T}_\mathrm{A}$\xspace}
\newcommand {\topoB}{$\mathrm{T}_\mathrm{B}$\xspace}
\newcommand {\topoC}{$\mathrm{T}_\mathrm{C}$\xspace}

\begin{document}

\title{Identifying edges that facilitate the generation of extreme events in networked dynamical systems}

\author{Timo Br\"ohl}
\email{timo.broehl@uni-bonn.de}
\affiliation{Department of Epileptology, University of Bonn Medical Centre, Venusberg Campus 1, 53127 Bonn, Germany}
\affiliation{Helmholtz Institute for Radiation and Nuclear Physics, University of Bonn, Nussallee 14--16, 53115 Bonn, Germany}

\author{Klaus Lehnertz}
\email{klaus.lehnertz@ukbonn.de}
\affiliation{Department of Epileptology, University of Bonn Medical Centre, Venusberg Campus 1, 53127 Bonn, Germany}
\affiliation{Helmholtz Institute for Radiation and Nuclear Physics, University of Bonn, Nussallee 14--16, 53115 Bonn, Germany}
\affiliation{Interdisciplinary Center for Complex Systems, University of Bonn, Br{\"u}hler Stra\ss{}e 7, 53175 Bonn, Germany}

\begin{abstract}
The collective dynamics of complex networks of FitzHugh-Nagumo units exhibits rare and recurrent events of high amplitude (extreme events) that are preceded by so-called proto-events during which a certain fraction of the units become excited.
Although it is well known that a sufficiently large fraction of excited units is required to turn a proto-event into an extreme event, it is not yet clear how the other units are being recruited into the final generation of an extreme event. 
Addressing this question and mimicking typical experimental situations, we investigate the centrality of edges in time-dependent interaction networks. We derived these networks from time series of the units' dynamics employing a widely used bivariate analysis technique.
Using our recently proposed edge centrality concepts together with an edge-based network decomposition technique, we observe that the recruitment is primarily facilitated by sets of certain edges that have no equivalent in the underlying topology.
Our finding might aid to improve the understanding of generation of extreme events in natural networked dynamical systems.
\end{abstract}
\maketitle

\begin{quotation}
Many natural, technological, or social systems are capable of recurrently generating large events that can lead to disasters when interacting with exposed or vulnerable human or natural systems. 
The understanding of the dynamical underpinnings of the generation of such extreme events has recently attracted much attention. 
While certain dynamical mechanisms have already been identified, only little is know about potential pathways in networked dynamical systems, that may play a vital role in facilitating the build-up of precursor structures that eventually lead to an extreme event.
We here use the concept of centrality --~originally proposed in the social sciences for network vertices and recently extended for network edges~-- to identify such pathways in networks of coupled, weakly interacting nonlinear oscillators. These networks are prototypical for excitable systems and are capable of self-generating and self-terminating extreme events.
We demonstrate that particularly interactions and only rarely edges in the coupling topology facilitate the build-up of precursor structures of extreme events. 
\end{quotation}

\section{Introduction}
Extreme weather events and other natural hazards, large-scale blackouts in power supply networks, market crashes, mass panics, wars, harmful algal blooms in marine ecosystems, or epileptic seizures in the human brain are recurrent, large-impact events that occur spontaneously in many natural, technological or social dynamical systems~\cite{hobsbawm1994,bunde2002,sornette2003,albeverio2006,ghil2011,mcphillips2018,farazmand2019}.
For systems that can be described by a time-dependent (or evolving) interaction network, novel methods have been developed over the last years that allow one to identify precursors of extreme events~\cite{havlin2012}. 
This holds true particularly for climate extremes~\cite{malik2012analysis,boers2013complex,scarsoglio2013climate,boers2014prediction,ludescher2014very,marwan2015complex,weimer2015predicting,gelbrecht2017complex,ozturk2018complex,boers2019complex,dijkstra2019application,ozturk2019network}, seismic extremes~\cite{duenas2007seismic}, hydrological extremes~\cite{konapala2017review}, economic extremes~\cite{constantin2018network,bosma2019too,faggini2019crises}, and epileptic seizures~\cite{lehnertz2016,rings2019precursors}.
Methods employed so far either aim at assessing global networks properties (e.g., clustering-coefficient-related or path-related measures) or local network properties --~mostly vertex centralities~\cite{Lue2016}. 
For interaction networks --~in which an edge represents attributes of an interaction (strength, direction, coupling function) between two vertices~-- an improved characterization of edge properties could add to advance understanding, prediction, and control of such networks~\cite{Slotine2012}.
To this end, and in order to find which edges in a network are important between other pairs of vertices, we recently modified various, widely used centrality concepts for vertices to those for edges~\cite{brohl2019centrality}.
We also proposed a network decomposition technique that is based on edge centrality and allows one to identify a hierarchy of sets of edges, with each set being associated with a different level of importance~\cite{brohl2019centrality}.

We here apply these novel concepts to investigate precursor structures of extreme events in the dynamics of complex networks of excitable units of FitzHugh–Nagumo type. 
Previous studies~\cite{ansmann2013,karnatak2014,bialonski2015,ansmann2016,rings2017} have shown these systems to be capable of self-generating and self-terminating strong, rare, short-lasting, and recurrent deviations from their regular dynamics without the influence of noise or parameter change.
These extreme events are preceded by local excitations (so-called proto-events~\cite{ansmann2013,rings2017}) in a certain fraction of units that play a decisive role in their generation.
Similar phenomena were also observed in other excitable systems~\cite{alvarez2017predictability,bonatto2017extreme,jin2017generation,karnatak2017early,kingston2017extreme,saha2017extreme,mishra2018dragon,saha2018riddled,moitra2019,ray2019intermittent}.
It is, however, not yet clear how the other units in a network are being recruited into the final generation of an extreme event, and we here conjecture that the recruitment is facilitated by certain edges.
We demonstrate the suitability of our novel concepts for the analysis of empirical data by mimicking typical experimental situations.

\section{Methods}

\subsection{Networks of excitable units}
We consider networks of $V$ diffusively coupled, excitable FitzHugh-Nagumo units $(n \in\left\{1,\ldots,V\right\})$, where the equations of motion of unit $n$ read   
\begin{eqnarray} \label{eq:fitzhughCoupled}
\dot{x}_n &=&x_n (a_n-x_n) (x_n-1)-y_n \\ \nonumber
&+& \frac{\cpl}{V-1}\sum_{m=1}^{V} A_{nm} (x_m-x_n)\\ \nonumber
\dot{y}_n &=&b_n x_n - c_n y_n.
\end{eqnarray}
The unit's internal control parameters are $a_n$, $b_n$, and $c_n$, and the coupling strength is denoted by \cpl.  
The symmetric adjacency matrix $\bfA \in \left\{0,1\right\}^{V \times V}$ has entries $A_{nm} = A_{nm} = 1$, iff units $n$ and $m$ are coupled.
We employ parameter settings that had been identified in previous studies~\cite{ansmann2013,karnatak2014,rings2017} to allow robust generation of extreme events in complex networks.
In particular, we set parameters $a$ and $c$ identical for all units: $a_n=a=0.0274 \, \forall n$ and $c_n=c=0.018 \, \forall n$; 
the parameter $b$ is mismatched with $b_n=0.006 + \frac{n}{V-1} 0.008, \forall n$,
and the coupling strength \cpl is chosen individually for each network.
We regard three coupling topologies each of which connects $V=20$ vertices but with different number of edges $E$: 
\begin{itemize}
\item{\topoA:} a binary network with a small-world topology~\cite{watts1998} with $E=100$ (number of nearest neighbors: $5$; rewiring probability: $0.25$) and $\cpl=0.128$;
\item{\topoB:} a binary network with a small-world topology~\cite{watts1998} with $E=40$ (number of nearest neighbors: $2$; rewiring probability: $0.25$) and $\cpl=0.33$;
\item{\topoC:} a binary network with a scale-free topology~\cite{barabasi1999} with $E=36$ and $\cpl=0.1128$. The degree ($\kappa$) distribution $\mathcal{F}$ of the network follows a power law of the form $\mathcal{F}(\kappa)\propto\kappa^{-3}$.	
\end{itemize}
Each networks' dynamics was integrated using an adaptive, explicit Runge-Kutta method of 5th order~\cite{ansmann2018} with a step size of $11$.
We discarded at least $10^4$ initial time units, and time series (here: $x$-components) used for further analyses consisted of $10^6$ data points.
The choice of the initial conditions (near the attractor) had no influence on our observations. 

\begin{figure}[htbp]
	\centering
	\includegraphics[width=0.5\textwidth]{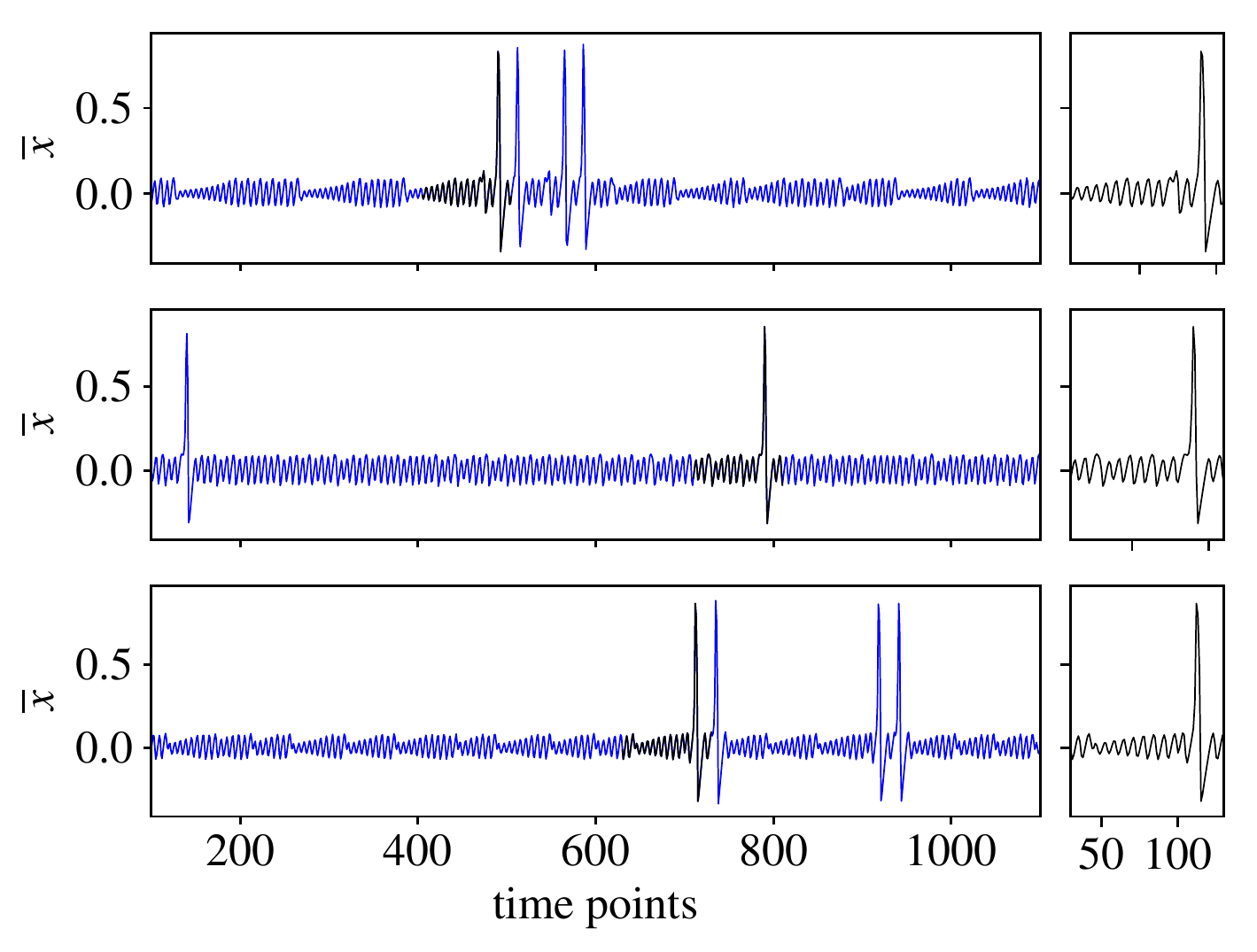}
	\caption{(left) Exemplary temporal evolutions of the average of the first dynamical variable $\overline{x}$ for topologies \topoA~--~\topoC (top to bottom). (right) Excerpt around an extreme event (colored black) of the respective time series.}
	\label{fig:examples}
\end{figure}

In Fig.~\ref{fig:examples} we show, for each coupling topology, excerpts of the time series of the average of the first dynamical variable $\overline{x}(t)=\frac{1}{V}\sum_{n}^{V}x_n(t)$.
Generally, we observe $\overline{x}(t)$ to exhibit irregular, low-amplitude oscillations~\cite{ansmann2013,karnatak2014,rings2017} with $-0.15<\overline{x}(t)<0.15$.
Occasionally, we observe stereotyped events at which all units become excited and thus $\overline{x}(t)$ clearly exceeds --~by at least a factor of six~-- the amplitude of the collective low-amplitude oscillations.
We consider these rare but recurring high-amplitude events as extreme events (time interval beginning with $\overline{x}(t)$ exceeding a threshold $\theta=0.5$).
We find 195 such events for \topoA, 138 events for \topoB, and 830 events for \topoC. 
For \topoA and \topoC, for which we often observe double extreme events~\cite{ansmann2013,rings2017}, only the leading one is considered.

\subsection{Data-driven construction of time-dependent interaction networks}
Mimicking typical experimental situations~\cite{bialonski2010,lehnertz2014,gao2017complex,brugere2018}, we derive time-dependent interaction networks by estimating --~using a sliding-window approach~-- the strength of interaction between pairs $\left\{n,m\right\} \in \left\{1,\ldots,V\right\}$ of time series of the first dynamical variables $x$. 
To do so, we employed an established method for investigating time-variant changes in phase synchronization.
The mean phase coherence~\cite{mormann2000} is defined as
\begin{equation}
    \label{eq:R}	
R_{nm}=\left|\frac{1}{T}\sum_{t=0}^{T-1}{e^{\iu\left(\Phi_n(t)-\Phi_m(t)\right)}}\right|,
\end{equation}
where $\Phi_n$ are the instantaneous phases of time series from unit $n$ (we here use the Hilbert transform~\cite{mormann2000,boashash1992_book}), and $T$ denotes the number of data points. 
By definition, $R_{nm}$ is confined to the interval [0,1], where $R_{nm}=1$ indicates fully phase-synchronized units.
Note that the window size $T$ is a critical parameter since it affects the sensitivity of the mean phase coherence.
We here chose $T$ such that a window captured at least one full cycle of either an extreme event or a low-amplitude oscillation.

Having calculated $R_{nm}$ for all pairs $(n,m)$ of units, we derived --~for each window~-- a synchronization matrix whose non-diagonal elements were associated with an adjacency matrix \bftA.
This matrix represents an undirected, weighted snapshot network.
To simplify notation, in the following we define $\tilde{A}_{nn} \defi 0 \,\forall\, n$.
Depending on the underlying coupling topology we refer to these time-dependent interaction networks as networks A, B, and C.

We position a reference window (window number $0$) around the extreme event, such that the window center coincides with the first time point for which the amplitude of $\overline{x}(t)$ exceeds the threshold $\theta$.
The window number increases while going back in time, with time windows of number $6$ or larger are assumed to represent typical inter-event dynamics.

\subsection{Estimating edge importance in time-dependent interaction networks}
\label{sec:centralities}
For our investigations, we employ two opposing~\cite{brohl2019centrality} concepts to estimate the centrality of edges in each snapshot network, namely edge betweenness centrality \CBE and edge eigenvector centrality \CEE. 

Edge betweenness centrality (of edge $k$) can be defined as~\cite{freeman1977,girvan2002}
\begin{equation}
\mathcal{C}^{\rm B}_{\rm e}(k)=\frac{2}{V(V-1)}\sum_{i\neq j}\frac{q_{ij}(k)}{G_{ij}}, 
\label{eq:CBE}
\end{equation}
where $k \in \left\{1,\ldots,E\right\}$, $\left\{i,j\right\} \in \left\{1,\ldots,V\right\}$, $q_{ij}(k)$ is the number of shortest paths between vertices $i$ and $j$ running through edge $k$, and $G_{ij}$ is the total number of shortest paths between vertices $i$ and $j$. 
A shortest path is defined as the path between two edges for which the sum of the inverse weights of edges along this path is minimal~\cite{brohl2019centrality}.

Edge eigenvector centrality (of edge $k$) is defined~\cite{brohl2019centrality} as the $k$th entry of
the eigenvector~$\vec{v}$ corresponding to the dominant eigenvalue $\lambda_{\max}$ of matrix ${\bf M}$, which we derive from the eigenvector equation ${\bf M} \vec{v} = \lambda \vec{v}$ using the power iteration method:
\begin{equation}
\mathcal{C}^{\rm E}_{\rm e}(k)=\frac{1}{\lambda_{\max}}\sum_{l}^{}M_{kl}\,\mathcal{C}^{\rm E}_{\rm e}(l),
\label{eq:CEE}
\end{equation} 
with $\left\{k,l\right\} \in \left\{1,\ldots,E\right\}$.
Here ${\bf M}$ denotes the edge adjacency matrix $\bftA^{\rm (e)} \in \mathbb{R}_+^{E \times E}$ whose entries $\bftA^{\rm (e)}_{ij}$ are assigned the average weight of edges $i$ and $j$ if these edges are connected to a same vertex, and 0 otherwise. 

With the aforementioned definitions, we regard an edge with the highest centrality value as most important (rank 1) and the one with the lowest centrality value as least important (rank $E$). 
In case of equal centrality values we rank in order of appearance.

\subsection{Identifying important sets of edges in time-dependent interaction networks}
With the aforementioned edge centrality concepts, we employ our previously proposed edge-centrality-based network decomposition technique~\cite{brohl2019centrality} that allows us to identify a bottom-up hierarchy of sets of edges (or ``webs''), where each set is associated with a different level of importance.
The decomposition technique consists of the following steps:
\begin{enumerate}[start=0]
\itemsep0em
\item initialize algorithm: set $E'=E$ and set iteration $q=1$;
\item estimate centrality $\mathcal{C}_{\rm e}(k)$ for all edges $k\in \left\{1,\ldots,E'\right\}$ in the current network;
\item choose the lowest centrality value as threshold value $\Theta=\min_k\mathcal{C}_{\rm e}(k)$, in order to eliminate less central edges;
\item every edge $k'$ with $\mathcal{C}_{\rm e}(k')\leq\Theta$ is assigned to the web of rank $q$ and is removed from the current network (which decreases $E'$; note that the $<$ sign holds for repetitions of step 3 within the $q$th iteration);
\item repeat step 1 and step 3 until no further edge is assigned to the web of rank $q$;
\item continue with next iteration (increase $q$ by 1) at step 1, as long as there are remaining edges to be assigned to webs;
\item reverse ranking of webs; the most important web has rank 1. 
\end{enumerate}
We note that this network decomposition can lead to two divisions of a network that are not helpful in identifying sets of edges associated with different levels of importance.
These cases are either an assignment of all edges to only one web (number of webs $\NWebs = 1$) or an assignment of each edge to a web ($\NWebs = E$). 
We also note that edges in a web do not have to be connected with each other.

\section{Results}
\begin{figure}[htbp]
	\centering
	\includegraphics[width=0.49\textwidth]{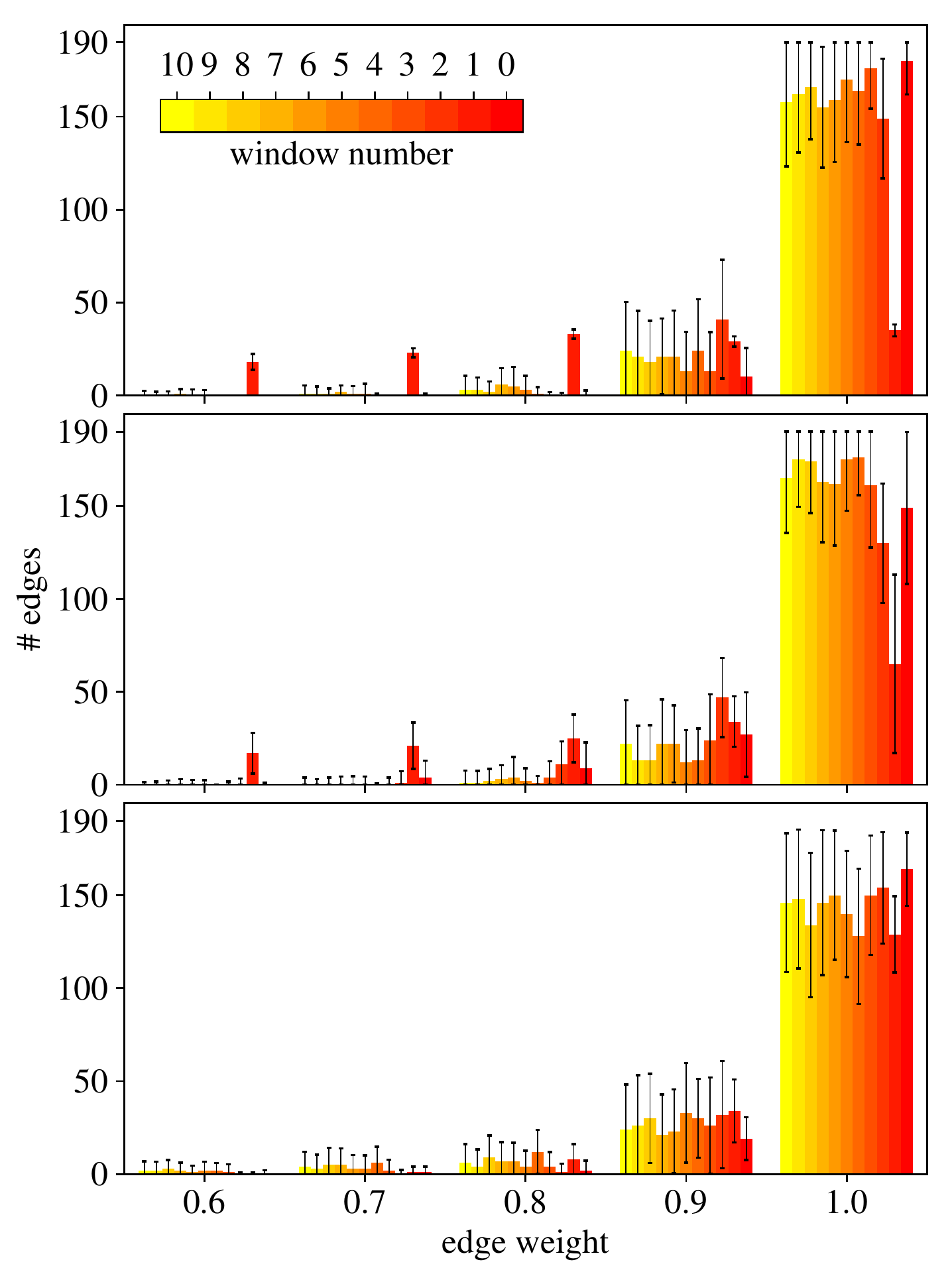}
	\caption{Edge weight distributions (means and standard deviations obtained from observations of the respective amount of extreme events) of time-dependent interaction networks A, B, and C (from top to bottom) for each time window. Time window $0$ is positioned around the extreme events, and the window number increases while going back in time.}
	\label{fig:weightDist_FULL}
\end{figure}

As shown earlier~\cite{ansmann2013,rings2017}, extreme events in the dynamics of coupled FitzHugh-Nagumo oscillators are preceded by proto-events during which a fraction of the units (those with low values of the control parameter $b$) become excited and which turn into extreme events, if and only if this fraction is sufficiently large (note that not all proto-events are followed by an extreme event).
It is, however, not yet clear how the other units are being recruited into the final generation of an extreme event.
We conjecture that the recruitment is facilitated by certain edges (or sets thereof), and in the following, we will identify and characterize these edges employing the edge-centrality concepts and the edge-based network decomposition technique.
Given that most of the complexity of a interaction network is encoded into the topology of interactions among its vertices (i.e., edges) and into the layout of the interactions' weights~\cite{barrat2004b,park2004,vanMieghem2005,Yang2009}, we first investigate how the edge weight distributions change when our time-dependent interaction networks transit into an extreme event.
Since edge weights represent the strengths of interaction between units (estimated with $R_{nm}$; see Eq.~\ref{eq:R}), 
we expect a narrow range of large edge weights despite the constant and rather low coupling strengths \cpl.
Indeed, the edge weight distribution peaks close to the maximum value of $R_{nm}=1$ with a rather narrow spread, by construction (see Fig.~\ref{fig:weightDist_FULL}).
\begin{figure*}[htbp]
	\centering
	\includegraphics[width=\textwidth]{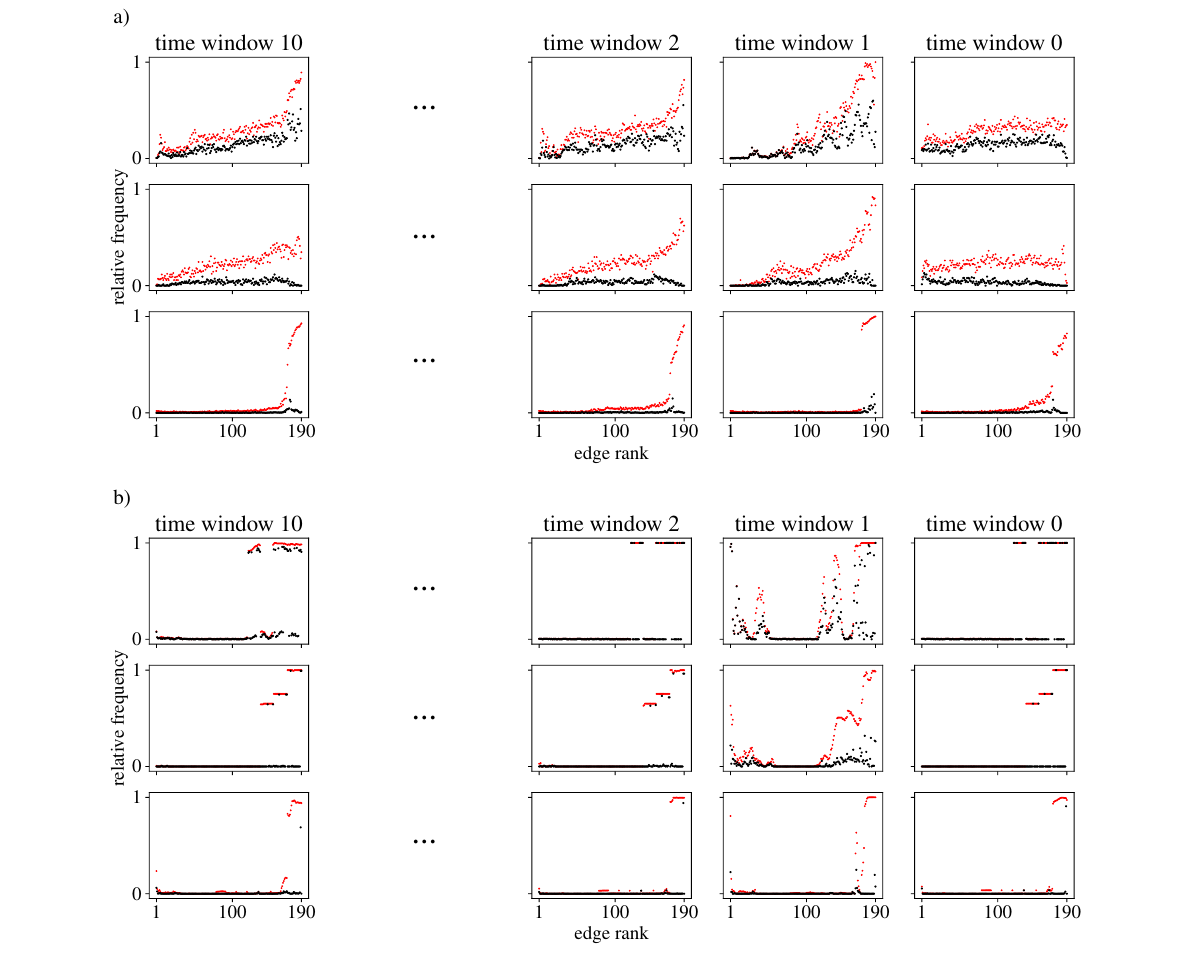}
\caption{Relative frequency of an edge with a given rank to be connected to a vertex whose dynamics exhibits proto-events. 
Edge rank estimated via ranking of a) edge eigenvector centrality and b) edge betweenness centrality.
Time window $0$ is positioned around the extreme events, and the window number increases while going back in time. Data from 195, 138 and 830 extreme events in the time-dependent interaction networks A, B, and C (from top to bottom). Red dots indicate edges in the time-depended interaction networks and black dots edges from the underlying coupling topology.}
	\label{fig:protoEdge}
\end{figure*}
For our time-dependent interaction networks, we find edge weights from the time window capturing an extreme event to compare to those from most preceding windows.
Interestingly though, we observe decreased edge weights in the time window directly preceding the extreme event (window 1), and this decrease is most pronounced for networks A and B.
With our analysis approach, proto-events thus reflect a desynchronized state during which only few units are simultaneously excited, while the other units are not.
We note that similar desynchronization phenomena were observed prior to epileptic seizures recorded in humans~\cite{mormann2003a} and in a simple dynamical model of two interacting networks of integrate-and-fire neurons that mimics such an event~\cite{feldt2007}.

Since edge weights impact on the centrality concepts employed here (cf. Sec.~\ref{sec:centralities}) and given our previous observations, we next hypothesize that a certain amount of edges in the interaction networks from the time windows prior to the extreme event will rank among the ones with highest centrality and are therefore possibly more relevant for the recruitment of further units.
In order to check this hypothesis, we estimate --~for each time window~-- the probability $\mathcal{P}$ for an edge to be identified as most important (i.e., highest centrality value and thus highest rank) with the respective edge centrality.
For each network, and independently of the used centrality, we observe (data not shown) in each time window (including time window 0) the respective probability distributions to peak around a small amount of edges (if we neglect edges with $\mathcal{P}<0.2$).
In addition, we observe that these distributions differ in the time window prior to the extreme event, indicating that during this time window other edges are most important.

Given these findings, we further investigate which edges are connected to vertices whose dynamics exhibit proto-events and whether these edges have a high rank and can be traced back to the underlying coupling topology (direct edge) or not (indirect edge).
As shown in Fig.~\ref{fig:protoEdge}, we observe edges with low rank to be (on average) more frequently connected to such vertices in all time windows preceding time window 1.
In time window 1, we additionally observe few more high-ranked edges to be frequently connected to these vertices, however, this findings holds for importance estimated using \CBE only.
If importance was estimated using \CEE, the low-ranked edges are even more frequently connected to these vertices.
Whereas the underlying coupling topologies had no influence on these findings, the differences seen for the two edge centralities can be related to the differences in their conceptual basis.
While \CEE considers the centrality of the neighborhood of a given edge, \CBE is a path-based approach to identify a central edge.
For either centrality concept it is rather straightforward to understand that in time windows far from the extreme event, high-ranked edges are not expected to be connected to the few vertices whose dynamics exhibit proto-events.
In the time window prior to the extreme event, the opposite can be observed with \CBE. 
This indicates that the recruitment of non-excited vertices is facilitated via short paths from excited vertices, making it more likely for edges that are directly connected to such vertices to have a high rank.
On the other hand, with \CEE highest-ranked edges connect non-excited vertices as these are mostly stronger connected (larger edge weights) to other non-excited vertices.

For direct edges and independent of their centrality ranking, we furthermore observe a general decrease, in time window 1 compared to other time windows, in their relative frequency to be connected to a vertex whose dynamics exhibits proto events.
One can thus deduce, that most of the edges that are connected to a vertex whose dynamics exhibits a proto-event represent indirect edges, with few exceptions found with betweenness centrality.

\begin{figure}
\includegraphics[width=0.5\textwidth]{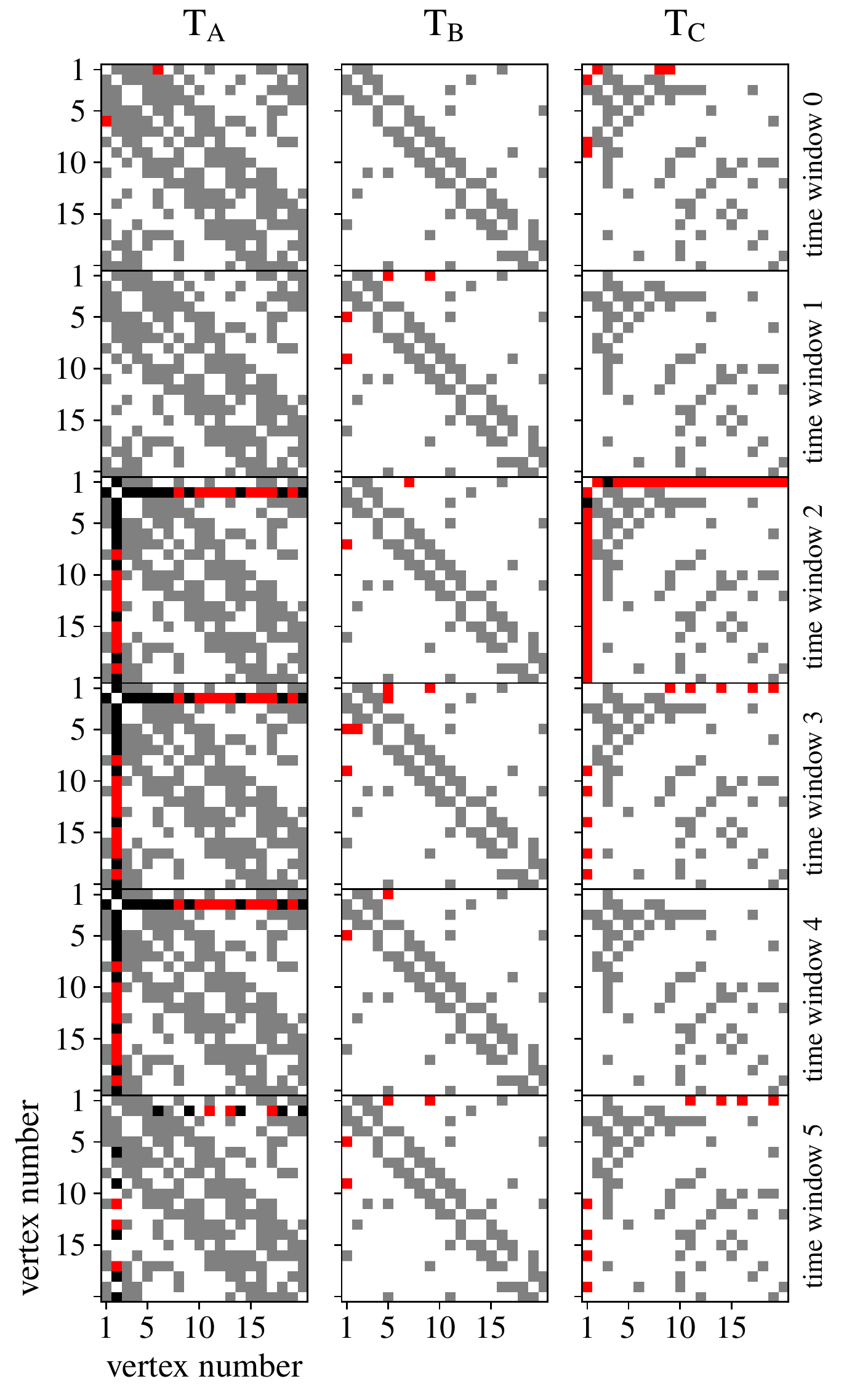}
\caption{Adjacency matrices of least important webs projected onto the underlying coupling topology (grey) with direct and indirect edges marked black and red, respectively. 
Only edges with a relative frequency (occurrence in data from respective amount of extreme events) higher than 75 \% were considered. 
\CEE-based network decomposition. Vertices sorted in ascending order of the control parameter $b$.}
\label{fig:impMatrixE}
\end{figure}

\begin{figure}
\includegraphics[width=0.5\textwidth]{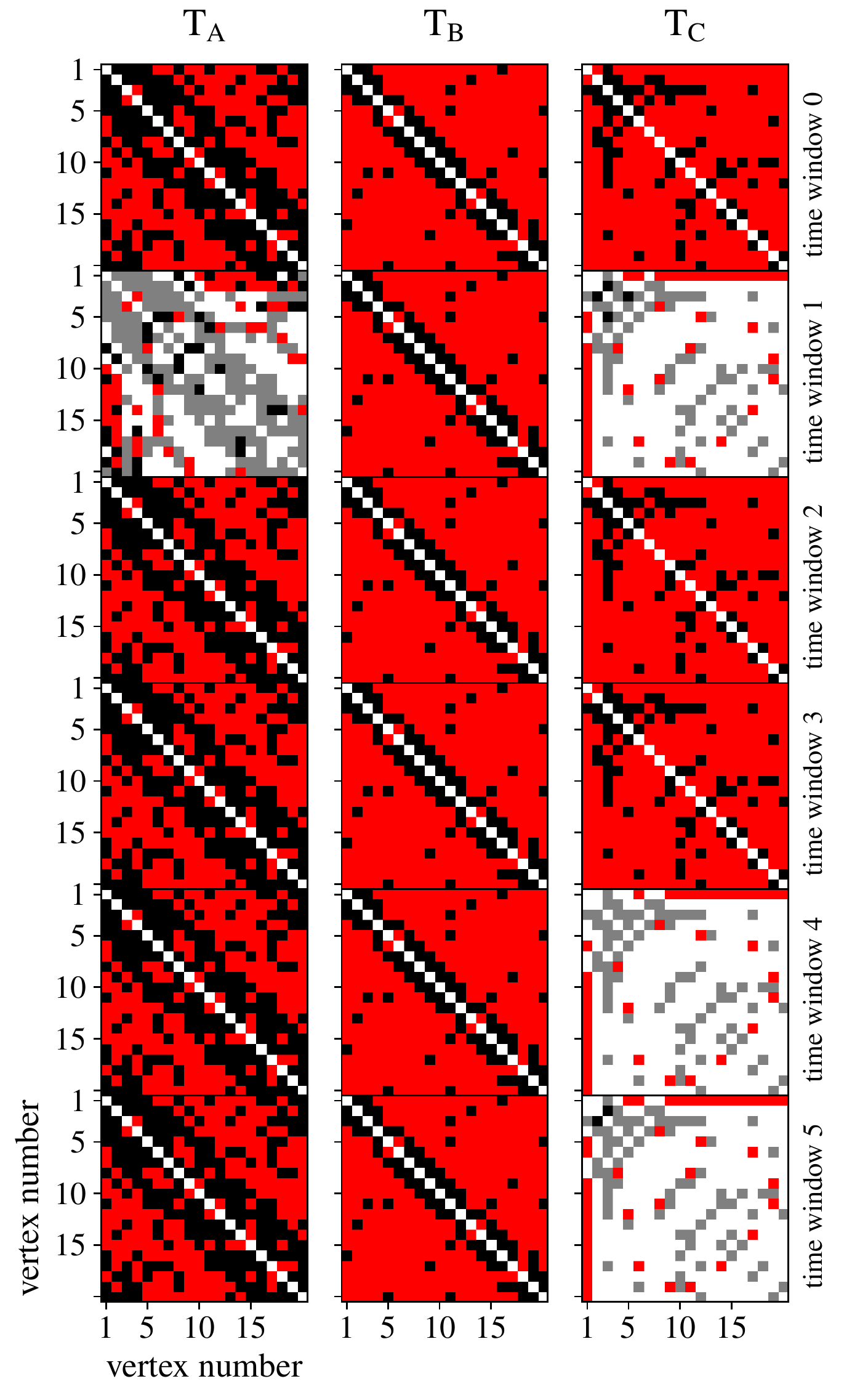}
\caption{Same as Fig.~\ref{fig:impMatrixE} but for \CBE-based network decomposition.}
\label{fig:impMatrixB}
\end{figure}

Summarizing our findings discussed so far, we conclude that the recruitment of non-excited units into the generation of an extreme event is facilitated by the most important and the least important  indirect edges.
As a last point, we investigate whether these edges form specific sets.
To this end, we employ our edge-centrality-based network decomposition technique to identify --~for each time window~-- the most and the least important web and eventually detail their characteristics. 
Our results presented in Figs.~\ref{fig:impMatrixE}~and~\ref{fig:impMatrixB} indicate that the least important webs for time window 1 consist of smaller sets of edges than the ones in the least important webs identified for windows preceding window 1 or even for the window that captures the extreme event. 
These sets consists to a greater amount of indirect edges than of direct edges.
In general, differences are most distinct for networks A and C and for the \CBE-based network decomposition.
We note that we achieved similar findings when considering the most important webs (data not shown).

Interestingly, the sets seen for time window 1 are composed of edges (either direct or indirect ones) connected to vertices whose dynamics exhibit proto-events. 
As expected the \CBE-based decomposition mostly identifies long-range connections while the \CEE-based decomposition mostly identifies  (nearest or next-nearest) neighboring connections within the web. 
It can be summarized that distinct sets of primarily indirect edges appear to play a vital role (in the time window prior to the extreme event) for the recruitment of non-excited units into excitation leading up to an extreme event. 

\section{Conclusions}
\label{sec:Conclusion}
We investigated which edges in networks of coupled, excitable FitzHugh-Nagumo units facilitate the recruitment of non-excited units into the final generation of an extreme event. 
With an eye on typical experimental situations that explore excitable system, we investigated the importance of edges in time-dependent interaction networks. We derived these networks from investigating the strength of interaction between time series of the units' dynamics in a time-resolved manner.
Importance of edges and sets thereof were characterized with the concept of edge centrality and an edge-centrality-based network decomposition technique respectively.
Our findings indicate, that the recruitment of non-excited units is facilitated primarily by sets of certain most and least important edges, both of which have no equivalent in the underlying topology.
A more comprehensive understanding of the role of such indirect edges and their relationship to the underlying coupling topology might aid to gain further insights into the generation of extreme events in natural networked dynamical systems.

\section*{Acknowledgements}
The authors would like to thank Lina Zabawa and Thorsten Rings for interesting discussions and for critical comments on earlier versions of the manuscript. 
This work was supported by the Deutsche Forschungsgemeinschaft (Grant No: LE 660/7-1).

\section*{Data availability}
The data that support the findings of this study are available from the corresponding author upon reasonable request.

\section*{References}

\end{document}